\documentclass[preprint,authoryear]{elsarticle}
\pdfoutput=1
\usepackage{times}
\usepackage[brazil,english]{babel}
\usepackage[T1]{fontenc}
\usepackage{epsfig}
\usepackage{graphics}
\usepackage{graphicx,url}
\usepackage{color}
\usepackage{amsfonts}
\usepackage[bookmarks=true,pdftex,colorlinks,bookmarksopen,bookmarksnumbered,citecolor=black,urlcolor=blue]{hyperref}
\usepackage{mwlabinputs}
\usepackage{paralist}
\biboptions{square} 

\begin{document}

\begin{frontmatter}

\title{Reflection-based language support for the heterogeneous capture and 
restoration of running computations\tnoteref{t1}}
\tnotetext[t1]{This work has been partially supported by CNPq Brasil}

\author[dcc]{Anolan Milanés\corref{cor1}}
\ead{anolan@dcc.ufmg.br}
\author[puc]{Noemi Rodriguez}
\author[puc]{Roberto Ierusalimschy}

\cortext[cor1]{Corresponding author}
\address[dcc]{Department of Computer Science,\\
Universidade Federal de Minas Gerais (UFMG)\\
Av. Antônio Carlos, 6627 - Prédio do ICEx - Pampulha - CEP: 31270-010,\\
Belo Horizonte - Minas Gerais - Brasil }
\address[puc]{Department of Computer Science,\\
  Pontifícia Universidade Católica do Rio de Janeiro (PUC-RIO)\\
  Rua Marquês de São Vicente, 225, Gávea \\
  Rio de Janeiro, RJ -- Brasil -- 22453-900}

\begin{abstract}
This work is devoted to the study of the problem of
user-level capture and restoration of running computations
in heterogeneous environments. 
Support for those operations has traditionally been offered through
ready-made solutions for specific applications, which are
difficult to tailor or adapt to different needs.
We believe that a more promising approach would be to
build specific solutions as needed, over a more general framework
for capture and restoration.
In this work, in order to explore the basic mechanisms a language 
should provide to support the implementation of different policies,
we extend the Lua programming language with an API that 
allows the programmer to reify the internal structures of execution into 
fine-grained language values.
\end{abstract}

\end{frontmatter}

\section{Introduction}

Systems with support for migration, persistence, monitoring, or debugging,
all have to deal with the non-trivial problem of capturing and restoring
the execution state.
It would thus be natural to expect a fair amount of reuse or sharing
among the developers of such systems.
However, decisions regarding issues such as what should be captured,
at what moment, and what to do in case of errors, vary a lot depending
on the intended application.
Traditionally, decisions about these issues are ingrained in systems
with support for migration or persistence, hampering the possibilities
of reusing them for purposes even slightly different from the original ones.

To avoid the need of building new systems from scratch every time
the requirement for state capture and restoration arises,
it would be interesting to factor out the support for capture and restoration
from other system decisions,
providing a mechanism flexible enough to acommodate different uses.
If a programming language can offer these mechanisms, it can then
be used to build libraries with the appropriate decisions (policies)
for different application areas.

It is natural do turn to reification and reflection 
in our search for  basic mechanisms for state capture and restoration.
{\em Reification} mechanisms allow execution state information to be made available
to the programmer as first-class values,
and {\em reflection} allows the programmer to alter program state
by incorporating data from the program into the execution state~\citep{FW84}.
However, reification can be supported at different granularities;
in~\citep{FW84}, for instance, the focus is on reifying program state by creating
continuations, which contain the complete ``remaining'' execution.
In order to support different migration and persistence strategies,
we believe it is necessary to provide fine-grained reification
and reflection mechanisms,
allowing the programmer to code decisions about what, when and
how to capture and restore.

In this paper, we discuss the motivation for this approach
and experiment with it by presenting an extension to the Lua programming language~\citep{Ierusalimschy06},
called {\em LuaNua},
with mechanisms for fine-grained \textit{reification} and \textit{installation}
of program state.
Lua has a number of reflective facilities which already provide
partial support for reification and installation, and Lua's design
relies heavily on the idea of separating mechanisms from policies,

The rest of this paper is organized as follows. 
Section \ref{approach} discusses the motivation of this work, 
as well as topics related to the design of state manipulation mechanisms  and our 
approach. Section \ref{sec:related} provides an overview of the state of the art on flexible support for heterogeneous 
capture and restoration of the execution state of running computations. 
Section \ref{sec:LuaNua} proposes an API for capturing and restoring computations and explores its functionality  through some examples. 
Section \ref{sec:formalization} provides a formal definition for the operational semantics of the chosen API.
Section \ref{sec:experiments} describes some experiments we have developed to analyze the impact of 
the proposed API on performance.
We conclude on section \ref{sec:final} with some final remarks.

\section{A different approach is needed}
\label{approach}

Systems that support either migration, persistence, or monitoring and debugging of running computations, require 
the capability of capturing and restoring data related to executions. 
However, each of these areas presents different characteristics and requirements. 
Through a detailed study of systems that provide support for migration~\citep{MRS08},
we observed that, even in this specific category, there are many different requirements
according to the intended application.
We have identified the following list of questions that deserve
different answers according to the specific 
application intended for the mechanism of capture and restoration.

\begin{enumerate}

\item What part of the execution state should be saved and restored?

This is arguably the most important question in this list.
It encompasses the issues of
(i) granularity of capture, (ii) depth of capture, and
(iii) completeness of restoration. 

As regards granularity (i),
most existing systems with support for state capture and restoration
consider either {\em threads} or
{\em processes} as the basic unit of execution state.
However, many other alternatives may be useful.
In some cases, it may in fact be necessary to capture an entire application process;
this may be the case, for instance, when migration is being used for opportunistic
computing. 
A process that was created in a host machine may need to be transferred
to another host when a local user reclaims the machine.
It can even be the case that an entire virtual machine must be transferred to the new host.
On the other hand, in most cases migration does not dictate the transfer of the
entire computation.
In fact, this should be avoided, whenever possible, due to
efficiency and feasibility reasons.
In some cases, as in that of {\em remote evaluation}~\citep{FPV98},
only the part of the computation that is to be executed at the remote site needs
to be transferred; this may mean that only the current activation record must
be captured.
It is also possible that parts of the code, such as those provided by libraries,
are already available at every possible host, making it unnecessary
to process them.
In some other cases, parts of the execution state do not even make sense outside
the current execution environment.
In yet other situations, migration may be achieved by transferring only global
data. 
This is the case, for instance, in event-oriented programs:
because no stack-related information is maintained between the handling
of two events in an event loop, all state information
is kept in global variables.

The issue of depth of capture (ii) involves decisions
about the depth of the dependency-graph
to be visited when capturing a value.
For instance, when capturing a function that refers to global variables
and functions,
should these other values also be captured?

Providing the ability to fully restore computations (iii)
implies capturing informations such as the program counter and
execution stack, 
generating additional overhead.
In many cases, however, there is either no need for complete execution restoration,
or it can be achieved by combining initialization data with code
(for instance, in the trivial case when the stack is empty).

\item How should errors be handled?

What should be done when, for instance, some system-dependent
information is found while traversing the object graph?
System-dependent data, such as file or socket descriptors, typically
cannot be simply captured and restored at another point in time or
space, because it will probably make no sense.
Even if the platforms are the same, descriptors are often simply
numbers that do not make sense outside the process in which they are
created; in heteogeneous environments, the problem is even more
complex.

If the goal of capture and restoration is to migrate computations in
opportunistic computing, input and output files will typically be taken along (with
or without the help of virtualization) and will need to be reopened at
restoration.
In other cases, such as that of remote evaluation or mobile agents, it
may be the case that only local
resources will be used at the new host, and thus the system does not
need to worry about
orchestrating the transfer of descriptors.

Handling errors depends on the requirements of the specific system
under construction.
This suggests that the programmer should have means to decide how to
treat those
situations, instead of being provided with ready-made solutions.

\item Should restored computations be bound to an existing execution context or 
should restoration create a new context?

In some applications, the obvious answer is to create a new process to
execute the restored computation.
This happens, for example, when migrating an application that is not
inherently distributed, and is being moved for the sake of load balancing.
In other cases, as in parallel programming, it may make sense to integrate
the restored computation in a running process (for instance, to avoid 
inter-process communication costs).


\item When should execution state be captured?
(and who should determine that it is to be captured?)

In many applications of migration and persistence, there is a notion that
state should only be captured when the computation is in a consistent state.
However, a generic notion of consistency is not easy to define.
For instance, if a single thread of a running application is to be
migrated, it may be necessary that state capture occurs outside 
any chunk that contains conceptually atomic updates of global variables.
The issue is easier to address when capture is triggered by the computation itself.
When the procedure is initiated from outside the computation, it is
complicated to determine that the consistency criteria are satisfied.
Besides, issues of authorization also arise.

\end{enumerate}

At a certain design stage, developers of systems and languages that deal with capture
and restoration~\citep{LS99, AF02, MRS08}
usually take decisions about the questions we listed.
These decisions simplify the use of the system for specific cases but affect 
their range of application.
While the resulting system fits well for the application that initially motivated it, maintaining
or extending the application after a certain amount of time can be complicated, specially when
unanticipated requirements appear. 
What if the amount of information grows to a limit that affects performance, creating the need
of controlling what to take along and what to rebind at the destination? 
The programmer may also try to use the system (s)he has already mastered
for a slightly different goal than the original one, but will probably find out that it
it will be necessary to
find and learn a new language that provides support for the new task at hand.

One solution that partially mitigates the problem, offering some flexibility,
is the provision of dump/undump-like calls to capture/restore a value, as in~\citep{TKS06}.
With these primitives, it is possible to implement both migration and persistence. 
However, these systems still contain ingrained decisions (such
as the depth of capture) that define the way they work.

As discussed in \citep{MRS08}, the problem is that current approaches are oriented to solve 
particular problems, instead of providing common support for the general case. 
To avoid the need of rebuilding from scratch every time a new problem arises,
languages and systems should be flexible enough to accommodate the needs of 
different users and applications, while allowing them to express their specific decisions. 
This can be achieved by offering generic mechanisms 
that allow the programmers
to implement their own policies in customized libraries and frameworks. 
These generic mechanisms must provide a common ground 
for capturing and restoring state,
upon which different applications can be  built.

We believe a good alternative for providing this common ground is to 
treat execution-state information as a set of fine-grained first-class values.
When the programmer captures such a fine-grained value,
(s)he should be able to easily inspect it to discover nested references
to other values and decide whether these should or not be captured.
Navigating down the captured values, s(he) can create
arbitrary data structures and use all available language features 
to code decisions about granularity, error-handling, and other issues.

What we are defending here is not that every application programmer should
face the task of composing his own support for persistence or migration
from fine-grained state manipulation primitives.
In fact, this is where libraries and the facility with which they are integrated into
a language come in.
The difference from traditional approaches is that (1) the resulting environments
are not specific, monolithic systems that do not interact with existing
libraries and frameworks; (2) the development of these domain-specific
libraries tends to be much easier and quicker than that of a monolithic
system.

In Milanés et al.~\citep{MRS08}, we discussed various proposals on heterogeneous migration of computations
and concluded that the lack of support for heterogeneous capture and restoration in most popular 
programming languages force them to resort to tricky methods, implying in penalties to performance, 
portability and maintainability.  
Most of the examined proposals were aimed at specific application areas.
There are works allowing both migration and persistence, but this support is often 
provided in the form of black-box mechanisms which follow a predetermined semantics. 
While capturing and restoring the execution state of a computation can be achieved straightforwardly with those \textit{dump}-like mechanisms, 
they limit the implementation of different policies. 

Languages with support for reflexion, like Java, allow capturing computations 
by constructing an object that contains all the state of the current executions. 
However, this requires the insertion of capture points across the user program, and also  
the insertion of code to be executed at restoration for restoring the state.  When it is required that the migrated/persisted program continue executing from the point it was suspended, the code must also restore the execution point where the capture was issued.
When the moment of the capture is unknown in advance (like in objective migration or in the case of 
persistence based on periodic checkpoints), the points of capture can be numerous, 
making it hard for the programmer to implement the required code for capturing and restoration for every possible suspension point. For instance, the code:
\begin{samplecode}
  obj.compute();
  obj.compute();
  obj.compute();
\end{samplecode}

basically will turn into:
\begin{samplecode}
  obj.compute();
  if (tocapture==true) { obj.save(1); }
  obj.compute();
  if (tocapture==true) { obj.save(2); }
  obj.compute();
\end{samplecode}

and the restoration into:
\begin{samplecode}
obj.restore(); 
switch(obj.entryPoint) {
     case 0: obj.compute(); 
     case 1: obj.compute(); 
     case 2: obj.compute(); 
}
\end{samplecode}

The ability of capturing computations is a main contribution of this work. 
Our proposal aims to facilitate the work of the programmer. The computation itself, including the stack and program counter instead of just object values, can be captured using high level library functions. It allows for  straightforward restoration, since the information of the point where the computation will be restarted is already encoded on the serialized data. Note that the fine-grained approach still makes possible to express specific behaviours, like diferentiate binding or how to handle non serializable values. 
Now the capture step is mostly done by a call to a library function and the restoration can be done by:
\begin{samplecode}
  obj.restore()
\end{samplecode}

It turns out that capturing computations by exploting this proposal allows for different behaviours that are difficult to express when following other approaches. For instance, there is a classification for migration methods according to the amount of execution state captured and restored, that divides them in weak and strong mechanisms. While weak mobility refers to the ability to allow code and, optionally, data transfer, strong mobility mechanisms allow migration of the execution state of a computation as well~\citep{MRS08}. Strong migration is desirable in various situations, such as in load-balancing applications for long-running, computation-intensive programs, and in the context of mobile agents. However, it is complex to build, and in consequence, is often implemented above weak mobility mechanisms. If we map functions to weak migration and computations to strong migration, we can see that our approach easies the implementation of both weak and strong mechanisms. 


\subsection{Proposal: A fine-grained reflective approach to capture and restoration} \label{sec:proposal} 

We have argued for handling
execution-state information as fine-grained first-class values.
We need mechanisms that allow the programmer to obtain these
first-class representations from the program state, and
symmetric mechanisms for installing these program-level
representations as part of the program state.
Friedman and Wand~\citep{FW84}  referred to these processes, respectively, as
{\em reification} and {\em reflection}.
Other authors have used the term {\em reflection} to define 
the processes of querying execution state information, reifying it into data
structures, and modifying the execution state~\citep{Smith82}.
To avoid ambiguities, we will refer to the processes that
Friedman and Wand called {\em reification} and {\em reflection}
as, respectively, {\em reification} and {\em installation}.
Those are the main operations provided by the API we propose.

Reification and installation are symmetric operations.
The reification mechanism must generate representations
that allow the programmer to choose exactly which items (s)he wishes to 
capture and to restore. 
Primitives should operate on fine-grained items, such as individual variables,
functions, and activation records.
The programmer should then be able to determine what other references are
contained in the resulting representations (for instance,
by accessing fields in an object), and thus control the 
level of recursion in which (s)he desires capture or
restoration to occur.
It is up to the  programmer to define the degree of similarity the 
representation and the represented data would have, since not all data may be 
of interest or even possible to install. 

Values returned by the reification mechanism should be copies,
or snapshots, of the represented entities. 
Changes to either the reified representation or to the execution
state itself should not affect each other.


In our approach, primitives for state capture return language values.
If, as often is the case, there is the need to produce byte strings with state representation,
this is carried out in a second step, that of {\em serialization}.
Conversely, a \textit{deserialization} step allows to convert a string of bytes
back into a program-level representation, which can then be installed into
the program state.
{\em Serialization} and {\em deserialization} procedures are supported
in many languages~\citep{Ierusalimschy06, RWWB96} and are not discussed in this paper.

We extended the programming language Lua~\citep{Ierusalimschy06,IFC07} with support
for fine-grained reification and installation.
We implemented primitives for \textit{reification} 
of values (capture) and for their \textit{installation} (restoration)
as an extension of Lua 5.1 that we have called \textit{LuaNua} 
(portuguese for ``Naked Lua''). 
Lua offers a number of reflexive mechanisms and
emphasizes the separation of mechanisms and policies by facilitating
language extensions~\citep{ry96}.

\section{Related Work}
\label{sec:related}

Lua already allows capturing and restoring Lua values, except for computations (in Lua, coroutines). 
However, these operations are incomplete 
and produce representations that are not easy to inspect.
On the other hand, Lua allows for easy integration of third party modules, and there is 
already a third party library for the serialization/deserialization of Lua values, called Pluto \citep{Sunshine-Hill08}. 
Pluto allows serializing any Lua value, and it is possible to provide a customized serialization procedure. 
It has, however, disadvantages compared to our approach: (i) it is not possible to restore sharing correctly if 
values are captured independently, (ii) as in Lua serialization, the representation generated is a bytestring 
(iii) policy in case of errors is fixed: if errors are detected (like no-serializable values), the procedure stops, 
(iv) as any third party software, it needs maintenance following the versions of the language. 

Languages with support for serializable continuations facilitates capturing and restorating computations. 
This is the case of Gambit-C~\cite{Feeley}, an implementation of Scheme that features the serialization of closures and continuations, as well as platform independence for the code representation. 
To deal with non-serializable objects, Gambit-C allows the programmer to define custom serialization and deserialization methods~\citep{GFM06}. 

Many others have, like us,  explored reflection for capturing and restoring computations. 
Iguana~\citep{GC96}, Reflex~\citep{TNCC03} and Geppetto~\citep{RDT08} are object-oriented extensions based 
on pre-processing code (C++, Java and Squeak, respectively) in order to add metalevel modifications, 
allowing the reification of various features of the language. 
Iguana and Reflex are not based on dynamic languages, thus the code transformations that allow reifying 
executions need to be done at load-time. 
The idea is to reify entities and operations to the metalevel.
For instance, it is possible to intercept a message from a particular class in order to execute some operation. 
Reflection in class-based object-oriented languages is mainly structural, that is, the objective is to create a reflective object model, but not a reflective model of the execution. 
This is different from our work  in that we reify the computation itself as 
inspectable entities and in that we propose integrating reflective facilities for 
capture and restoration into the language itself.

Programmer intervention in serialization is a controversial issue for 
being so laborious and error prone. 
However, it has supporters both in the context of persistence and distributed systems. 
Sewell et al.\citep{SLW+07} exploit this approach for controlling the interaction 
between instances of different versions of the same program coexisting in a 
distributed system, in the context of the Acute programming language.
In Acute, programmer intervention consists in giving 
hints to the compiler. For instance, (s)he can insert marks to indicate limits for marshalling, 
so, event though \textit{thunkification} (serialization of a computation) is an atomic operation,
it is possible to control its extension. 
Likewise, in the context of persistent applications, manual intervention has been studied as a solution for maintenance and versioning. 
The project of the E programming language~\citep{Miller06} put a strong emphasis on that issue. Indeed, they argue that manual persistence based on a schema made by the programmer is the right solution to better deal with the ``schema evolution'' problem~\citep{Miller} (This problem refers to how to upgrade an application without loss of previous user data). Manual persistence, in this case, allows minimizing the information to be saved, thus the problem can even be avoided in some cases, or at least, overcomed. Like LuaNua, \textit{E} follows the tradition of mechanism/policy separation by offering a set of building blocks on top of which persistent systems can be built.

Java is frequently mentioned in discussions about serialization.
In Java, persistence can be achieved through object serialization~\citep{RWWB96}. 
Java serialization represents objects as stream of bytes. 
Besides, serialization is deep, and there is no way to specify which subset 
of the data will be serialized. 
Java computations (ie. threads) are not serializable. 
Object-relational mapping libraries for Java simplify the implementation of persistence by 
overcoming the mistmatch between the Java object model and the relational database. 
However, thread persistence is not provided. 

\section{LuaNua: Fine-grained reification and installation in Lua}\label{sec:LuaNua}

In this section, we describe the extensions we introduced in Lua
for fine grained capture and restoration.
We first present a brief description of Lua, then describe the API,
and finally discuss some examples of its use.

\subsection{A brief intro to Lua}

Lua  is an interpreted, procedural and dynamically-typed language. 
It is based on prototypes and features garbage collection. 
Lua values can be of type \textit{nil}, \textit{boolean}, \textit{number}, \textit{string}, 
\textit{table}, \textit{function}, \textit{thread}, and \textit{userdata}. 
Tables are the language's single data structuring mechanism and
implement associative arrays, indexed by any value of the language except \textit{nil}. 
The type \textit{thread} is used for Lua coroutines.
Coroutines are lines of execution  with their own stack and instruction pointer,
sharing global data with other coroutines.
In constrast to traditional threads (for instance, posix threads), coroutines
are collaborative: a running coroutine suspends execution only
when it explicitly requests to do so.
Closures and coroutines are first-class values in Lua. 

Lua supports the reification of functions as strings of bytes.
However, this representation is not easily handled, and 
may require translation in case of different architectures.
The language does not allow for the serialization of coroutines.

In Lua, source code and bytecode chunks can be loaded and executed dynamically.
Other reflective features include access to the environment and the names of local variables.
The fact of being an interpreted language ensures portability for the code.
The simplicity of concurrency in Lua avoids the need to address problems of synchronization.
Lua coroutines are \textit{stackful}, meaning they can suspend (and restart) execution 
at an arbitrary level of function calls.
Having coroutines as first class values allows an homogeneous treatment of data
and computations.

Finally, Lua offers a large part of its functionality through libraries. 
LuaNua is basically an extension of the set of reflexive resources of Lua 
with \textit{reify} and \textit{install} operations.

\subsection{LuaNua}

The LuaNua extension adds very few functions to Lua:
basically \textit{reify} and \textit{install}:
\begin {itemize}
  \item \textit{reify(value, [level])} receives a value as parameter, and returns the representation 
of its structure. 
This primitive accepts an optional second parameter, which makes sense only in the case of 
reification of coroutines and represents the level of the desired activation record.
  \item \textit{install(representation, type | value,[level])} receives two parameters: the representation 
and the type or the value to be rebuilt. 
If successful, an invocation of this function should return a value of the type specified.
Like \textit{reify}, \textit{install} receives an aditional parameter ``level'' 
for the installation of coroutines. 
\end {itemize}

The API also offers two generic auxiliary functions: \textit{name} and \textit{fields}.
\textit{name} returns a unique identifier for structured values (here the uniqueness is guaranteed only on the platform of execution) for identity. This is necessary to maintain sharing on restoration and to identify values that are being received back in their original setting. 
Function \textit{fields} returns the description of the representation of any kind, for documentation. The only argument is the name of the type.

Representations returned by \textit{reify} contain only atomic values 
(numbers, booleans, strings), and references to structured abstractions 
(tables, functions, upvalues, prototypes, threads, userdata). 
Values such as tables and functions are reified only at their first level, and can be navigated
by the programmer who can then decide whether or not to reify inner values.
In the specific case of coroutines, 
extraction and installation are made on the level of the activation record, 
thus allowing control over how many levels are handled. 
Installation requires the reconstruction from the inside to the outside of the values.

Structured values in LuaNua are reified as Lua tables.
Because tables can be indexed by any type of value and can grow
as needed, this facilitates construction and traversal of state
representation.
Reification of arbitrary extents of program state  with LuaNua 
thus consists of the progressive construction of a representation of 
the execution, by traversing the tables returned by successive calls to
{\em reify} and filling them in, as needed, with the results of new invocations.
The resulting tables can then be serialized using standard language 
mechanisms~\citep{Ierusalimschy06}.

\textit{reify} must return a copy of this representation 
of the value instead of providing direct access to the structures of the execution. 
Any modification of this representation is made offline (as in StrongTalk~\citep{BBG+02}). 


It must be possible to modify current executions, not only create new computations. 
This is necessary in cases where only some coroutines of the application are modified, 
since the creation of a new coroutine creates a new reference that will not coincide 
with the value referenced by other entities within the program. 
For this reason, \textit{install} should allow receiving as a parameter the coroutine where 
the installation will be made, and should be able to modify or add the contents of any stack.

\subsection{Exploring the API}\label{sec:implementation}

In this section, we present a series of examples to illustrate
the  flexibility afforded by LuaNua.
The examples in this section  were executed saving the resulting state in a file 
and restoring the stored representation 
from this file in another instance of the Lua interpreter,
and thus would make sense in the context 
of both migration and persistence.

\subsubsection{Basics}\label{sec:basicex}

Our first  example shows how to reify and install a function using the proposed API. 
We choose a simple function (\textit{inc}) that receives a single parameter and 
returns its value incremented by one.
(The code presented is real Lua code. Comments in Lua begin with ``\--\--''; we are using the symbol \--\--> for output.) 
 \begin{samplecode}
  local function inc(counter)
    return counter + 1
  end
 \end{samplecode}

Using the LuaNua API, this function can be reified as follows:

\begin{samplecode}
-- reify function inc 
local tinc = debug.reify(inc)
print(tinc) --> {p = 0x532920}
\end{samplecode}

The call to {\em debug.reify} returns a table containing the 
representation of function {\em inc}, 
that is, a reference to the function prototype (and debug data not discussed 
here for simplicity). 
That reference is saved in field {\em p} of table {\em tinc} ({\em tinc.p}). 
Next, we proceed further into the representation by reifying the function prototype, 
so we make a further call to {\em debug.reify}:

\begin{samplecode}
local proto = debug.reify(tinc.p) 
\end{samplecode}

Now table \textit{proto} contains the bytecode of the function prototype. 
Because all the values composing the representation are already atomic, 
reification ends here. 

At this point, tables {\em proto} and {\em tinc} could be serialized and
written to a file or transferred in a message.
In any case, at some later point in time the tables could be reconstructed
and we would be ready to reinstall function {\em inc}.

For installation, we follow a bottom-up approach. 
First we install the internal, non-atomic, values in memory space. 
In this case, the only such value is the function prototype (type ``proto''), 
which was saved in table \textit{proto}. 
Thus, we install the function prototype (and save it as tinc.p) and 
then re-create (install) the function represented by table tinc. 

\begin{samplecode}
local tinc = {p = debug.install(proto,'proto')}
local newinc = debug.install(tinc,'function')
\end{samplecode}

Now that newinc contains the installed function, we can execute it:
\begin{samplecode}
print(newinc(1)) -->2
\end{samplecode}

We can see in this example that the reification/installation 
manual procedure allows the programmer to control the composition of the representation.
On the other hand, we see that values that were previously hidden 
(such as prototypes) are now visible. 
Because the language design does not assume that these values will be manipulated
by the programmer, this visibility may lead to unanticipated execution
errors (for instance, illegal operations on these values will not be handled
elegantly).

\subsubsection{Reification/installation of executions}

A more interesting example is capturing and restoring executing computations. 
Lua provides asymmetric coroutines, which are controlled through calls 
to the \textit{coroutine} module.
A coroutine is defined through an invocation of \textit{create} with an initial 
function as a parameter. 
The created coroutine can be (re)initiated by invoking \textit{resume}, 
and executes until it invokes \textit{yield}.
For instance, function \textit{count} is an iterador that, for each number from 1 to 5,
prints the number and yields its value
(in Lua, {\em coroutine.resume} returns a value of {\em true} or {\em false} indicating
whether the coroutine was resumed successfully and, optionally, values
passed to {\em yield}).:

\begin{samplecode} 
-- definição da função
local function count()
  for i = 1,5 do
    print("Number",i)
    -- send this number back to activator
    coroutine.yield(i)
  end
end
\end{samplecode}

Suppose we want to execute \textit{count}  until it produces the number 3, 
and then capture the suspended coroutine.
We can write the following code.

\begin{samplecode}
local coro = coroutine.create(count)
local status, i
repeat
  -- resume returns a status and
  -- the yielded values
  status, i = coroutine.resume(coro)
until (i == 3)
capture(coro)
\end{samplecode}
Let's turn our attention to the implementation of \textit{capture}.
Internally, a Lua coroutine has a stack organized in 
activation records, each of them corresponding to an active function. 
We can reify a Lua coroutine by composing its reified frames. 

We must iterate over the frames composing the stack, invoking
\textit{reify} for each level and unwinding every structured 
(that is, non atomic) value. 
This procedure creates a table with the representation of the requested 
entity and its components. 
(Later, this table can be converted into bytestrings for storage or transfer.).

We begin by obtaining a reference \textit{thr} to a new table:
\begin{samplecode}
local thr = {}
\end{samplecode}

Then we save the coroutine attributes, like its status:

\begin{samplecode}
thr.status = coroutine.status(coro)
\end{samplecode}
Now we iterate over the valid levels of the stack, from the top level:

\begin{samplecode}
repeat
  level = level + 1
  thr[level] = debug.reify(coro, level)
\end{samplecode}

Since coroutine representation contains non-atomic data, 
its contents must in turn be reified.
For that, we invoke \textit{save}, an auxiliary function
we constructed, which inspects the type of its argument 
and reifies it recursively if it is not atomic.
\textit{save} receives as arguments the value to be reified and a 
table of already reified values (the \textit{saved} table), 
so values are reified just once even if they appear many times.

To complete the loop, we increase \textit{level}, to move to
the next activation record:

\begin{samplecode}
repeat
  level = level + 1
  thr[level] = debug.reify(coro, level)
  thr[level] = save(thr[level],saved)
until (thr[level]==nil)
\end{samplecode}

To install an equivalent coroutine, we must load the representation 
saved in \textit{thr} and rebuild the coroutine structure. 

\begin{samplecode}
local ncoro = debug.newthread()

for i = #thr, 0, -1 do
  ncoro = debug.install(thr[i], ncoro, 0)
end

debug.setstatus(ncoro,thr.status)
print(coroutine.status(ncoro)) --> suspended
\end{samplecode}

Now we can resume our new coroutine and the next number in the iteration is printed:

\begin{samplecode}
coroutine.resume(ncoro) --> Numero 4
\end{samplecode}

If only a partial execution is desired, it can be constructed 
from scratch with a subset of the reified activation records.

\subsubsection{Controlling graph extension}
In order to improve performance and migratability, the programmer can omit non-essential data before transmission or persistence. 
This is the case, for instance, of variables whose content will no longer be used. 

As an example, consider the following chunk of code, in which a function receives 
a value from the standard input, processes it and prints the result. 
It then returns control to the caller.
\begin{samplecode}
local function myprint()
  local tofinish, a = true

  while tofinish do
    a = input_data()
    print(process_data(a))
    tofinish = coroutine.yield() --give another the chance to work
  end
end
\end{samplecode}

If the computation is captured when function \textit{myprint} has yielded, 
all data, including local variable \textit{a}, are part of its state, and will be serialized (note that \textit{a} could be a datum of any size). 
However, the value contained in \textit{a} has already been processed and will be thrown away, thus its current value can be replaced, for instance, with a null value.

\subsection{Pickling library}
\label{pickling}

It can be argued that
the degree of control offered by our approach introduces a burden
on the programmer. 
We expect the LuaNua API to be used not by final Lua programmers,
but by library developers who can implement specific policies.

As an example, 
we have extended the LOOP library~\citep{Maia} in order to facilitate capturing and 
restoring Lua computations. 
LOOP (Lua Object-Oriented Programming) is a set of packages that allows the implementation of 
different object-oriented programming models in Lua and offers limited support for serialization. 
Using the reification and installation mechanisms of LuaNua,
new functions were added for
capturing and restoring coroutines, functions, prototypes and upvalues. 

Serialization in LOOP
follows a deep approach and allows the serialization of values with self-references. 
The serialization algorithm is similar to the one described in \citep{TKS06}. 
It consists basically on generating expressions that describe the deserialization 
procedure for every value, 
while deserialization consists on the evaluation of those expressions. 
Values are serialized once and their ids are kept for future reuse. 
A value can only be installed after all the values it references are already installed: 
a recursive de-serialization on the structure of a value containing self-references 
can provoke an infinite loop. 
\textit{Promises} are created on de-serialization for such values. 
A promise is a reference to an empty value of the same type so further references to 
that value can use the promise instead of the proper value. 
When the value is ready to be installed, the promise is fulfilled. 

As an example, consider the function:
\begin{samplecode}
function func()
        return 4
end
\end{samplecode}

Serializing this function with LOOP will produce a string such as:
\begin{samplecode}
serial:setup(serial:value(0x5381e0,'function',
  serial:value(0x538300,'table',{
    ["p"]=serial:value(0x532f80,'proto',
      serial:value(0x5385e0,'table',{
        ["nups"]=0,["numparams"]=0,["sizek"]=1,
        ["k"]=serial:value(0x53ea70,'table',{[1]=4,}),
        ["code"]=serial:value(0x5386f0,'table',{[1]=1,[2]=16777246,[3]=8388638,}),
      })
    ),
  })
),nil,0,0x5381e0)
\end{samplecode}
which shows the contents of function \textit{func} and the contents of its prototype, 
namely, the number of its non-local variables (nups), parameters (numparams), 
number of constants (sizek), list of constants (k),
 and the bytecode (code). 
Information related to debugging is not shown here for simplicity. 

Our experience using the serialization framework shows that the size of the 
serialized string grows considerably with the complexity of the target value, 
which confirms our argument about the need to manipulate the data before transmission or persistence. 
Also, inserting calls to the serializer into third-party software demands some 
help from its developers, which makes sense given that, as we have argued, 
they know which is the really meaningful information. 

\subsubsection{Capture}
To capture and persist the state of the execution, our \textbf{capture} function issues 
a call to the instance of the LOOP serializer to process the  coroutine where 
the program was executed. 
As a result of this procedure, a chunk of code that allows re-creating the execution along with 
the environment can then be serialized.
\begin{samplecode}
function capture(co, stream)  
   stream:put(co)
   return string.format("
end
\end{samplecode}

After this, the returned string can be easily persisted.

Capturing files introduces the problem of how to capture non-portable values. 
To solve it, we added a new load method to our serializer library. 
It opens the file at filepath with the access mode and position it had when captured. 
Then the file is returned.

\begin{samplecode}
function restoreFile (self, filepath, mode, pos)
   if mode=='w' then mode='r+' end
   if mode=='wb' then mode='rb+' end
   local file = assert(io.open(filepath, mode))
   file:seek("set", pos)
   return file
end
\end{samplecode}

To be able to find and set the correct file access modes, we need to register 
that information when a file is opened. 
Only registered files will be captured, and their data must be stored at the serializer 
instance before serialization. 

\subsubsection{State restoration} 

State can be restored from the serialized value (\textit{buffer}) produced by the capture step. 
After loading the serializer libraries, we need to instantiate a serializer:
\begin{samplecode}
 local mystream = MyStream()
\end{samplecode}
Then we store the buffer in the serializer instance and call the deserialization method. 
The method returns the captured coroutine. 
\begin{samplecode}
 mystream.data=buffer
 local restoredCo = mystream:get()
 print(coroutine.status(restoredCo)) --> suspended
\end{samplecode}
Now the platform can resume execution:
\begin{samplecode} 
 local status, error = coroutine.resume(restoredCo) 
\end{samplecode}

\section{Operational semantics} \label{sec:formalization}

In this section, we specify the operational semantics of the reification and installation primitives 
of the proposed API. 
This semantics can be defined through operations executed on a machine that 
interprets a language formally specified on a particular representation of the state. 
Our definition is based on an extension of the SECD machine~\citep{Landin64} 
(chosen for its similarity to the Lua language) with support for assignment and multiple states. 
Henderson~\citep{Henderson80} offers an extensive description of the SECD machine 
and its instructions set. The notation used here is taken from~\citep{FF06}.

The SECD is a stack-based machine; the functions get their parameters from the stack. 
The state of the machine can be described through the contents of 4 registers 
(from where its name originates):
\begin{itemize}
  \item S (stack): stores the temporary results when it computes the value of expressions. 
Analogous to an activation record or frame;
  \item E (environment): stores the values bound to variables during the evaluation;
  \item C (control list): stores the bytecode of the program that is running;
  \item D (dump): used as a stack to store the values of the other registers when a new function is called.
\end{itemize}

The effect of an instruction can be defined through the states of these registers 
before and after the execution of the instruction.
\begin{figure}[hbt!]
\textit{C} = $\emptyset$ | b \textit{C} | x \textit{C} | ap \textit{C} | $prim_{o^n}$ \textit{C}\\
\textit{S} = $\emptyset$ | v \textit{S}\\
\textit{E} = is a function of variables to values $\{ \langle x,v \rangle, ... \}$\\
\textit{D} = $\emptyset$ | $\langle$ \textit{S, E, C, D} $\rangle$\\
v = b | $ \{ \langle { \langle x \; C \rangle } E \rangle \}$  
\caption{Language syntax}
\label{fig:syntax}
\end{figure}

The sets \textit{S}, \textit{E}, \textit{C} and \textit{D} are defined as shown in 
Figure~\ref{fig:syntax}, where b is a constant, x is a variable, ap is an application, 
$prim_{o^n}$ are n-arity primitive operations,
$\{ \langle { \langle x \; C \rangle } E \rangle \}$  are closures (pairs of open terms and environments E).
\newcommand{\<}{\ensuremath{\langle}}
\renewcommand{\>}{\ensuremath{\rangle}}
The relation $\longmapsto$ defines a one-step transition on the state of the machine SECD. 
The main transition rules on that machine are shown as follows:
\begin{eqnarray}
 \< S, E, b\; C, D\>
&\longmapsto& 
\< b\; S , E , C , D\> \label{regra:1}\\
\< S , E , x\; C , D\>  
&\longmapsto& 
\langle E(x)\; S , E , C , D\rangle \label{regra:2}\\ 
\langle b_n...b_1\; S ,  E ,  prim_{o^n}\; C ,  D\rangle
&\longmapsto& 
\langle \delta(o^n,b_1,b_2,...,b_n) \; S , E , C , D\rangle \label{regra:3}\\
\< S ,  E , \< x\; C'\> C , D \>
&\longmapsto&
 \<\<\< x\; C'\> E\> \; S ,  E ,  C ,  D\> \label{regra:4}\\
\< v\<\< x\; C'\rangle E'\rangle \; S,  E , ap\; C , D\rangle 
&\longmapsto& 
\<\emptyset , E'[x \leftarrow v] , C' , \< S, E, C, D \>\> \label{regra:5}\\
\< v\; S ,  E , \emptyset , \< S' , E' , C' , D \>\>
&\longmapsto&
 \< v\; S' ,  E' ,  C' ,  D\> \label{regra:6}
\end{eqnarray}

These rules state that:
\begin{description}
\item Rule \ref{regra:1}: the evaluation of a constant pushes the constant in the S register.
\item Rule \ref{regra:2}: the evaluation of a variable returns, in the S register, 
the value of the variable in the environment E.
\item Rule \ref{regra:3}: the application of a primitive operation on a list of values n 
returns a value in the S register.
\item Rule \ref{regra:4}: the evaluation of a $\lambda$ abstraction returns a closure.
\item Rule \ref{regra:5}: the application of a function on a parameter generates the 
creation of a new activation record in the stack and the storage of the previous values 
in the dump D. 
In the environment E, the parameter x is replaced by the value of the argument.
\item Rule \ref{regra:6}: At the end of the execution of a function call, 
the previous activation record is restored on the machine registers and the 
return value is pushed on the new stack.
\end{description}

We extend the machine SECD with assigments 
(Lua is a language with state) and multiple coroutines. 
A coroutine is represented by the registers SECD. 
To store these coroutines we added a register called  \textit{storage}($\Sigma$). 
The \textit{storage} is a function that maps locations to values.

To formalize the transfer of control, we need an additional register on the 
$SECD\Sigma$ machine that we shall call~\textit{activation stack} (A). 
Register A will store the active coroutines in order of activation: 
at the top of the list is the coroutine that is running at the moment. 
From now on, the top of A will be used as the registers of the machine, 
that is, we are defining the semantics of operations as transitions from $\< A, 
\Sigma \>$ into $\< A', \Sigma' \>$ where:

A = $\emptyset$ | $\langle \sigma, thr , A \rangle $  \\
where thr = $\langle S, E, C, D\rangle $ 

Thus, the definition of our configuration modifies and extends the definition in Figure~\ref{fig:syntax} with the sets:
\begin{description}
\item S = $\emptyset$ | v S | $\sigma$ S
\item E = a function of identifiers to locations $ \{\langle x, \sigma\rangle,... \} $
\item C = $\emptyset$ | b C | x C | ap C | $prim_{o^n}$ C | set C | newthread C | reify C | install C | resume C | yield C 
\item D = $\emptyset$ | $\< S, E, C, D \> $
\item v = b | $ \{ \langle { \< x \; C \> } E \rangle \}$ 
\item $\Sigma$ = a function that maps locations to values $\{\langle \sigma , v \rangle,... \}$
\item A = $\emptyset$ | $\< \sigma, \< S, E, C, D\> , A \> $
\end{description}

Following are the basic transition rules modified (*):
\small
\begin{eqnarray}
 \<\< \sigma, \langle S ,  E , b\; C , D \>, A \>, \Sigma\rangle
&\longmapsto&
\<\< \sigma, \langle b\; S , E , C , D \>, A \>, \Sigma\rangle \label{regra:n1}\\
\<\< \sigma, \langle S , E , x\; C , D \>, A \>, \Sigma\rangle
&\longmapsto& 
\<\< \sigma, \langle v\; S , E , C , D \>, A \>, \Sigma\rangle \label{regra:n2}\quad (*)\\ 
&&where\; v=\Sigma(E(x))\nonumber\\
\<\< \sigma, \langle b_n...b_1\; S ,  E ,  prim_{o^n}\; C ,  D \>, A \> , \Sigma\rangle
&\longmapsto& 
\<\< \sigma, \langle v\; S , E , C , D \>, A \>, \Sigma\rangle  \label{regra:n3}\\
&&where \; v=\delta(o^n,b_1,b_2,...,b_n) \nonumber\\
 \<\< \sigma, \langle S ,  E , \< x\; C'\> C , D \> , A \> , \Sigma\rangle
&\longmapsto& 
\<\< \sigma, \langle\langle\< x\; C'\> E\> S ,  E ,  C ,  D \> , A \> , \Sigma\rangle \label{regra:n4}\\
\<\< \sigma, \langle v \langle\langle x\; C'\rangle E'\rangle \; S,  E , ap\; C , D \> , A \> , \Sigma\rangle
&\longmapsto& \<\< \sigma, \langle\emptyset , E'[x \leftarrow \sigma'], C', \< S, E, C, D \> \>, A \>, \Sigma[\sigma' \leftarrow v] \rangle\nonumber\\
&& where \; \sigma' \notin dom(\Sigma) \quad  (*)\label{regra:n5}\\
\<\< \sigma, \langle v\; S ,  E , \emptyset , \< S', E', C', D \>  \> , A \> , \Sigma\rangle
&\longmapsto& 
\<\< \sigma, \langle v\; S' ,  E' ,  C' ,  D \> , A \>, \Sigma\rangle \label{regra:n6}\\
\<\< \sigma, \langle v\; S,  E, set\; x\; C, D \> , A \>, \Sigma\rangle
&\longmapsto& 
\<\< \sigma, \< v\; S, E, C,  D \> , A \> , \Sigma[\sigma' \leftarrow v]\rangle\label{regra:n7}\quad (*)\\
&&where \; \sigma'=E(x)\nonumber
\end{eqnarray}
\normalsize

The new rule \ref{regra:n7} defines assigment.

Now we define the rules that describe the semantics of the coroutines 
operators (\textit{create}, \textit{resume} and \textit{yield}). 

\small
\begin{eqnarray}
create\; \langle x\; C'\rangle E'::\nonumber\\
\<\< \sigma, \<\< x\; C'\rangle E'.S,  E,create.C, D \>,A \> , \Sigma\>
&\longmapsto& \<\< \sigma, \<\sigma'.S,E,C,D \>, A\>, \Sigma[\sigma'\leftarrow \< \< x\; C'\> E',\emptyset,\emptyset,\emptyset \>] \>\nonumber\\
&&where\; \sigma' \notin dom(\Sigma) \label{regra:coro1}\\
resume\; \sigma'\; v::\nonumber\\
\< \< \sigma, \< \sigma'\; v\; S,  E, resume\; C, D \> ,A \> , \Sigma \>
& \longmapsto
& \<\< \sigma' , \< v\; S',E',C',D' \>, \label{regra:resume}\\
&& \<\sigma ,\< S, E, C, D\> , A\> \> ,\Sigma [\sigma' \leftarrow nil]\>\nonumber\\
&& where \;\langle S',E',C',D'\rangle = \Sigma(\sigma')\nonumber\\
yield\; v::\nonumber\\
\<\<\sigma , \< v\; S , E , yield\; C , D\> , \;\;\;\;
& \longmapsto 
& \<\<\sigma', \< v\; S',E',C',D'\> , A \>, \label{regra:yield}\\
\<\sigma',\< S', E', C', D'\>, A \>\> ,\Sigma\> 
&& \Sigma[\sigma\leftarrow \< S,E,C,D\>]\> \nonumber\\
\nonumber\\
\<\<\sigma , \< v\; S,  E, \emptyset , \emptyset \>, \<\sigma' ,\< S',E',C',D'\>, A \>\>,  \Sigma\>
& \longmapsto & 
\<\<\sigma',\< v\; S',E',C',D' \> ,A \>, \Sigma \>\label{regra:coro4}
\end{eqnarray}
\normalsize

\begin{description}
\item Rule \ref{regra:coro1}: creating a coroutine consists in creating
a state and installing the closure $\< x\, C'\> E'$ 
on its first (and only) activation record.

\item Rule \ref{regra:resume}: (Re) starting a coroutine consists in moving it from 
the \textit{storage} to the top of A. The argument is placed at the top of S.

\item Rule \ref{regra:yield}: operation yield stores the current coroutine in 
the \textit{storage} and eliminates it from A. 
The previous stack is restored in the machine registers.

\item Rule \ref{regra:coro4} extends rule \ref{regra:n6} to 
describe the termination of a coroutine and the return of 
control to the coroutine that made the activation.
\end{description}

To formalize the reification and installation of coroutines we need lower 
level operations than those presented earlier. 
We can take advantage, however, of a function for installing closures which 
is embedded inside the operation for creating a coroutine 
(rule \ref{regra:coro1}). 
Thus, we decompose that operation into the creation of an empty coroutine 
and the installation of a closure on that coroutine:
\small
\begin{eqnarray}
newthread::\nonumber\\
\<\< \sigma, \< S,  E,  newthread\; C ,   D \>, A \> , \Sigma\> 
&\longmapsto& 
\<\< \sigma, \<\sigma'\; S, E, C , D \> , A \>, \Sigma[\sigma'\leftarrow\emptyset,\emptyset,\emptyset,\emptyset]\> \nonumber\qquad \\
&&where\; \sigma' \notin dom(\Sigma) \label{regra:newthread}\\
install\;\<\< x\; C'\> E'\>,\sigma'::\nonumber\\
\<\< \sigma, \<\<\<x\; c'\> E'\> \sigma' S,E,install\; C,D \>, A \>, \Sigma\> 
&\longmapsto& 
\<\< \sigma, \<\sigma'\; S, E, C, D \> , A \>, \label{regra:coro6}\\
&& \Sigma[\sigma'\leftarrow \< \< x\; C'\rangle E' \> ,\emptyset,\emptyset,\emptyset\>]\>\nonumber
\end{eqnarray}
\normalsize

Rule \ref{regra:newthread} describes the creation of a new empty coroutine and the 
return of the reference that represents it. 
The coroutine startup process ends with the insertion of the closure passed as parameter, 
on top of the stack (rule \ref{regra:coro6}). 
This is true both when startup occurs normally, through the interpreter, and when it is through operation 
\textit{install} of the proposed API. 
Actually, the install operation in rule \ref{regra:coro6} can be extended to the more general 
case of the installation of an activation record at any level, 
considering the record $\< \< \< x\; C'\> E \>, \emptyset, \emptyset, \emptyset \>$. 

The reification and installation of coroutines can be defined as follows:
\small
\begin{eqnarray}
reify\; \sigma', n::\nonumber\\
\<\< \sigma, \<\sigma' \; n\; S,E, reify\; C , D \>, A \>, \Sigma\>
&\longmapsto& \label{regra:content}\\
&&\<\< \sigma, \<\< S',E',C'\> S,E,C,D \>, A \>, \Sigma\>\nonumber\\
&&where\; \langle S',E',C',D'\rangle =  \nonumber\\
&&sublist(n,find(\sigma', \Sigma, A)) \nonumber\\
install\; \< S',E',C'\>,\sigma',n::\nonumber\\
\<\< \sigma, \langle \< S',E',C'\>\sigma'n S,  E, install\; C, D \> , A \> ,\Sigma \rangle
&\longmapsto& 
\<\< \sigma, \langle\sigma'S, E, C, D \>,A' \> , \Sigma' \>\label{regra:install}\\
&& where\; find(\sigma', \Sigma', A')= \nonumber\\
&& put(n, \< S',E',C'\>, find(\sigma', \Sigma, A))\nonumber
\end{eqnarray}
\normalsize
where \textit{sublist}, \textit{put} and \textit{find} are auxiliary functions to 
navigate through the stack in order to install and reify activation records.

\begin{description}
\item Rule \ref{regra:content}: Reification consists in the extraction of the SEC 
registers that corresponds to the requested frame. 
\item Rule \ref{regra:install}: Installing an activation record consists 
in copying it in the frame \textit{level} of the coroutine corresponding to reference $\sigma$.
\end{description}

Using the presented formalization we can prove that reification and installation are 
symmetrical operations, that is, 
\begin{eqnarray} 
	R(I(Rep)) & \equiv & Rep \nonumber\\
	I(R(V)) & \equiv & V\nonumber
\end{eqnarray}
which is to say that:
\begin{equation}
\<\< \sigma, \<\sigma' \; n\; \sigma'\;  n\;  S,E, reify\; install\; C , D \>, A \>, \Sigma\>
\longmapsto
\<\< \sigma, \<S,E, C , D \>, A \>, \Sigma\> 
\label{prova:show}
\end{equation}, and the inverse is also valid.

We show that no modification is produced when a reified record is reinstalled on the same level of the coroutine. 

From \ref{regra:content},  
\begin{eqnarray}
&&\<\< \sigma, \<\sigma' \; n\; \sigma'\;  n\;  S,E, reify\; install\; C , D \>, A \>, \Sigma\> 
\longmapsto\nonumber\\
&& \; \; \<\< \sigma, \<\<S',E',C'\> \; \sigma'\;  n\; S,E, install\; C , D \>, A \>, \Sigma\> \nonumber\\
&&where \nonumber\\
&&\<S',E',C',D'\> = sublist(n, \<S'',E'',C'',D''\>) ,\nonumber\\
&& \<S'',E'',C'',D''\> = find(\sigma', \Sigma, A) \nonumber
\end{eqnarray}
From \ref{regra:install} $\<\< \sigma, \<\<S',E',C'\> \; \sigma'\;  n\; S,E, install\; C , D \>, A \>, \Sigma\> 
\longmapsto\\
\<\< \sigma, \<\sigma'\; S,E,C ,D \>, A \>, \Sigma\>,$ \\
where $\sigma'$ references the coroutine $\<S''',E''',C''',D'''\> = 
put(n, \< S',E',C'\>, find(\sigma', \Sigma, A)) $
Thus, for \ref{prova:show} to hold, it is necessary that $\<S'',E'',C'',D''\> = \<S''',E''',C''',D'''\>$. The proof is a trivial induction on the level of the activation record being reified, for a stack with an arbitrary length. 

\section{Experiments} \label{sec:experiments}

To evaluate the overhead introduced by the reification and installation procedures, we 
implemented the capture and restoration of a program that calculates the factorial of a number, 
and the migration of the n$^{th}$ Fibonacci number and a k-NN (k-Nearest Neighbor) application. 
Those implementations use the pickling library described in Section~\ref{pickling}.
The experiments on capture and restoration where conducted to verify the behaviour of the 
checkpointing latency and the ratio between delays storage/load and capture/load. 
The capture time is defined as the time of reflecting the computation into a string 
(In LOOP the extraction of the representation of the computation as tables and its 
conversion to strings are performed together in the serialization functions, so we did not 
separate those times.). 
We measured the cost of the migration (migration latency) compared to the execution time 
of the computation, to verify if this migration procedure is reasonable.
The migration cost here is defined as the time between the begining of the capture and the end 
of the load stage (after the computation is installed back). 
The execution time of computations refers to the time elapsed between the start of the execution and 
the return of the function (with the final result). 

We also evaluated our proposal with a real 
application: a program executing the k-NN algorithm for the classification of a documents database. 

The experiments were executed on two CPUs Intel Core 2 Duo 2.16 GHz machines with 1 GB de RAM 
connected to a 100Mb switch inside a local network. 
Both machines run Fedora Core 8 kernel 2.6.25.4-10.  
Time is measured using the getrusage() function, thus it refers to the CPU time effectively used by the process. 
All time units are miliseconds.
The code for all examples is available at~\citep{LuaNua}.

\subsection{Capturing and restoring the Factorial recursive algorithm}

Because the capture and restoration times depend on the amount of data transferred, 
our first experiment measures the capture/storage and load/restoration times of a computation 
as the stack size increases.
We execute the following implementation of the Factorial recursive algorithm: 
\begin{samplecode}
 local function factorial(n)
    if n==0 then
        coroutine.yield()
        return 1
    else
        return n*factorial(n-1)
    end
 end
\end{samplecode}
and capture the coroutine in which the function is executing when a call to \textit{yield} is issued. 

\begin{figure}[ht]
\vspace{-.5cm}
\hspace{-.9cm}
\begin{minipage}[b]{0.6\textwidth}
\includegraphics[width=1.06\textwidth]{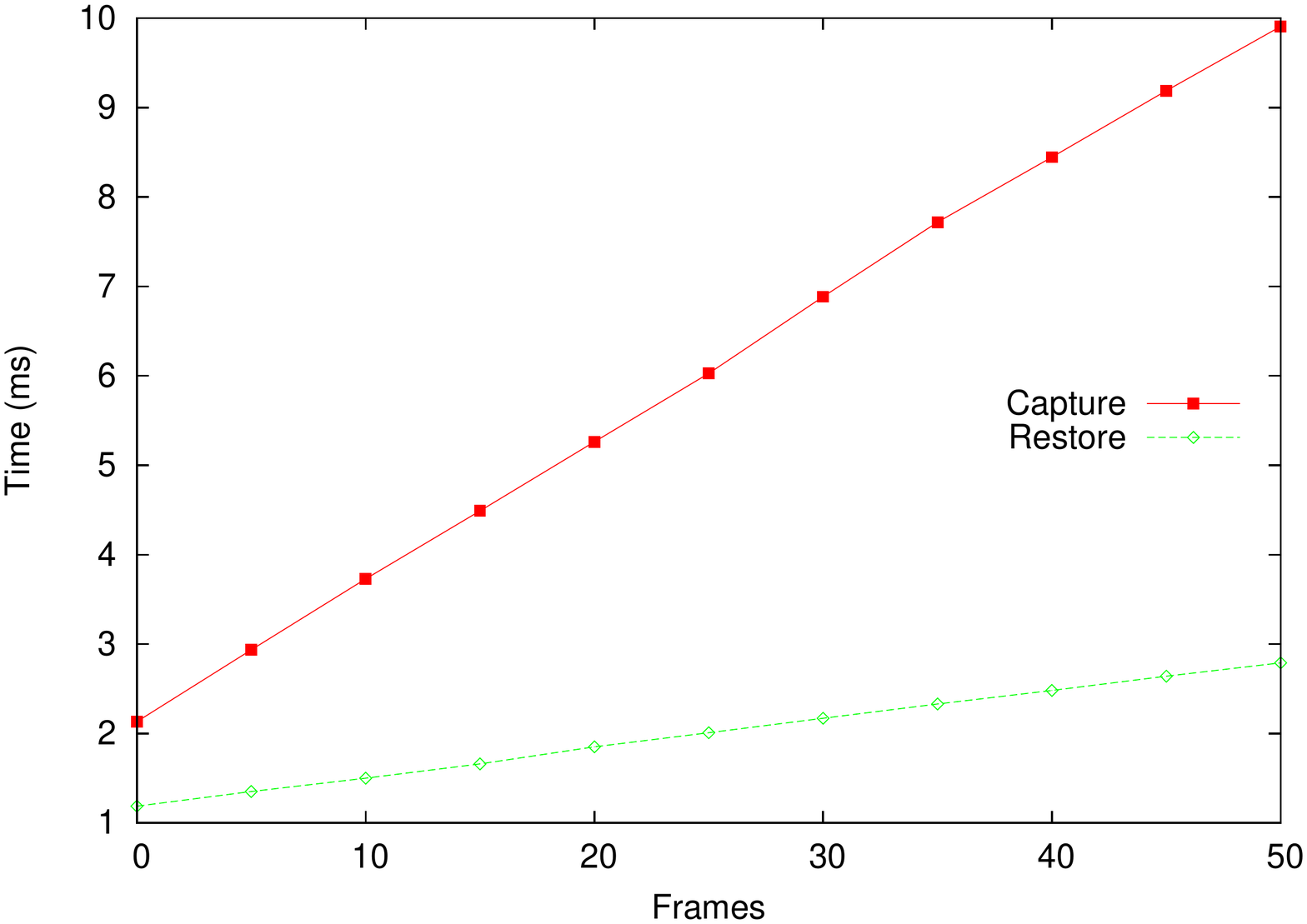}
\vspace{-1cm}
\caption{Capture/Restoration}
\label{fig:factCaptRest}
\end{minipage}
\hspace{-1.45cm}
\begin{minipage}[b]{0.6\textwidth}
\includegraphics[width=1.06\textwidth]{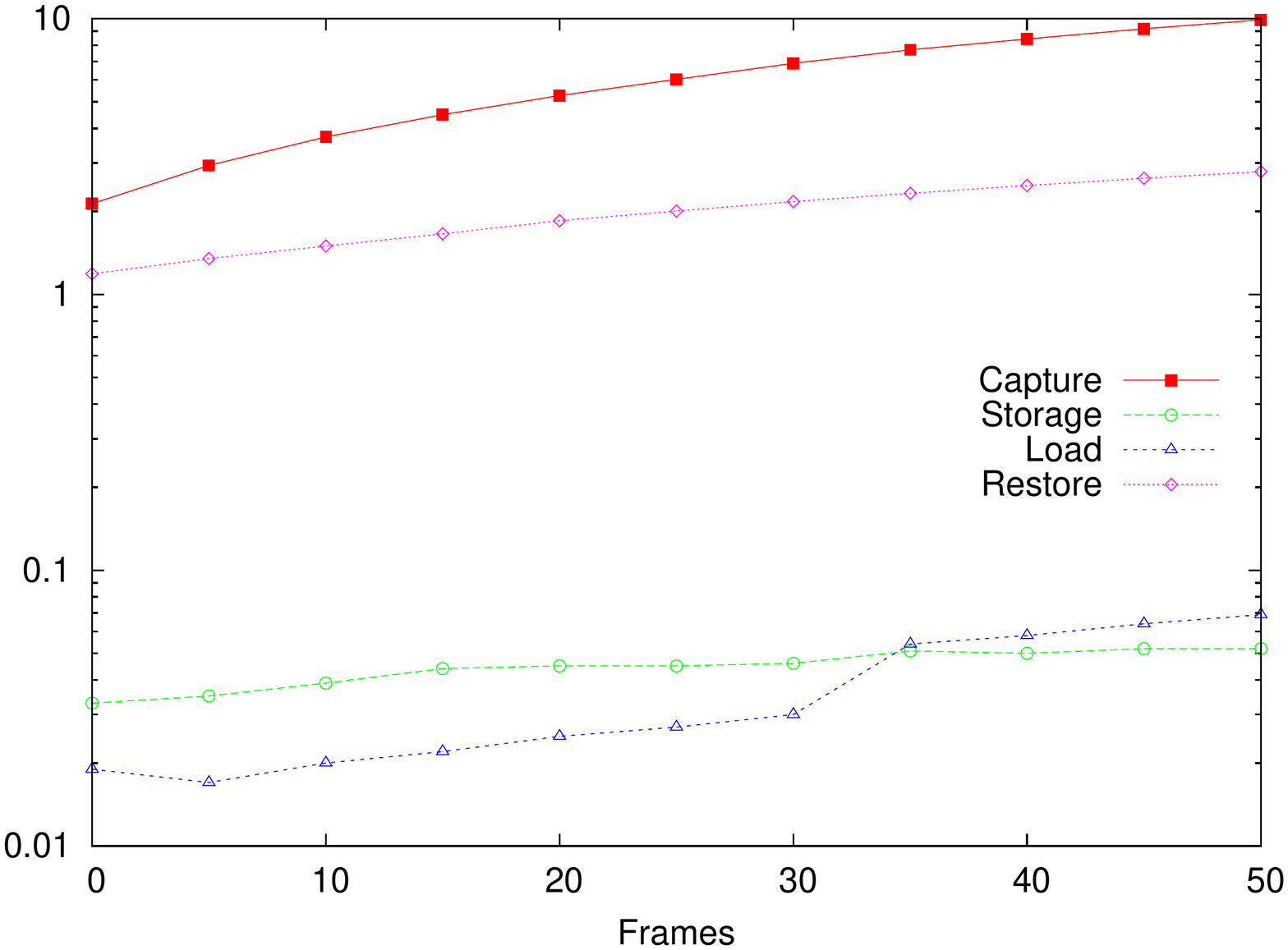}
\vspace{-1cm}
\caption{Capture/Store/Restoration/Load}
\label{fig:factTime}
\end{minipage}
\end{figure}

We ran this code for values from 0 to 50 (at intervals of 5).
Figure \ref{fig:factCaptRest} plots the delay of capture and restore operations 
we obtained for different frame sizes.
Because the storage and load time are very similar and close to zero, 
we used a logarithmic scale on the \textbf{y} axis of Figure \ref{fig:factTime} to plot 
the time for capturing, saving, loading, and restoring operations for each frame size.
 

The results show the higher complexity of capture when compared to restoration, 
which is due to the process of constructing the installation code
that will be executed at restoration. 
It is also apparent that both the capture and restoration times grow almost linearly 
with the number of frames on the stack, as expected, when the size of the data increases, 
while the delay for saving and loading keeps steadily negligible. 
That is, capture and restoration weigh far more than file-related procedures. 
Those results are rather different to other observations found in the literature(~\citep{BHKP+04}). 
This is partly due to the verboseness of our library and also to the nature of 
the reification/installation procedure, which is more devoted to flexibility than to performance. 
Unlike in Bouchenak's approach, in our case we are mapping an execution structure 
to a language value and then serializing the value. 

\subsection{Migration}

In this section, we discuss an experiment with migration.
We developed  a  migration platform based on luaNua, the pickling library discussed in
Section~\ref{pickling}, and ALua~\citep{URI02}, an event-based system for distributed programming.
Using this platform, we implemented two applications and measured migration times.
The first application is the Fibonacci recursive algorithm.
For the second application, we chose the k-NN algorithm because it is a real application, 
is easy to implement in Lua, has a regular behaviour and forced us to consider 
aspects not usually present in simpler experiments, such as dealing with open files. 

\subsubsection{Migration Platform}

The programmer must explicitly insert suspension points across his code
to allow migration to take place.
When the application yields and the platform regains control, 
the application is captured and transferred.
The captured computation is saved to a file that is copied to the
destination, as are any open files.
At the new host, the migration platform executes the received code
to recreate the computation, which is then restarted. 

To guarantee that all open files are transferred,
we redefined function \textit{io.open} to store information about opened files in a
table for latter use.

The experiments were performed using two machines but, to avoid dealing with clock differences,
the first machine (A) executes a computation, suspends it, captures its state in 
a file, sends it through the network to (B) (along with the opened files, it also sends a message 
for B to initiate the operation of sending them back), from where the files are resent to A and then restored. 
\textit{Migration time} was computed as the sum of capture, store, restoration and load time, 
plus the transmission time divided 2. 
The \textit{Total execution time} is the time elapsed between the initiation 
of the coroutine and
its return, including the time for migrating the coroutine from A to B. 

\subsubsection{Migration of a Fibonacci execution}

We executed the following implementation of the Fibonacci recursive algorithm: 
\begin{samplecode}
 local stop = true
 local function fibonacci(n)
    if n==0 then
        if stop then coroutine.yield(); stop = false; end
        return 0
    elseif n==1 then
        return 1
    else
        return fibonacci(n-1)+fibonacci(n-2)
    end
 end
\end{samplecode}

\begin{figure}[h!]
 \centering
\includegraphics[width=.9\textwidth]{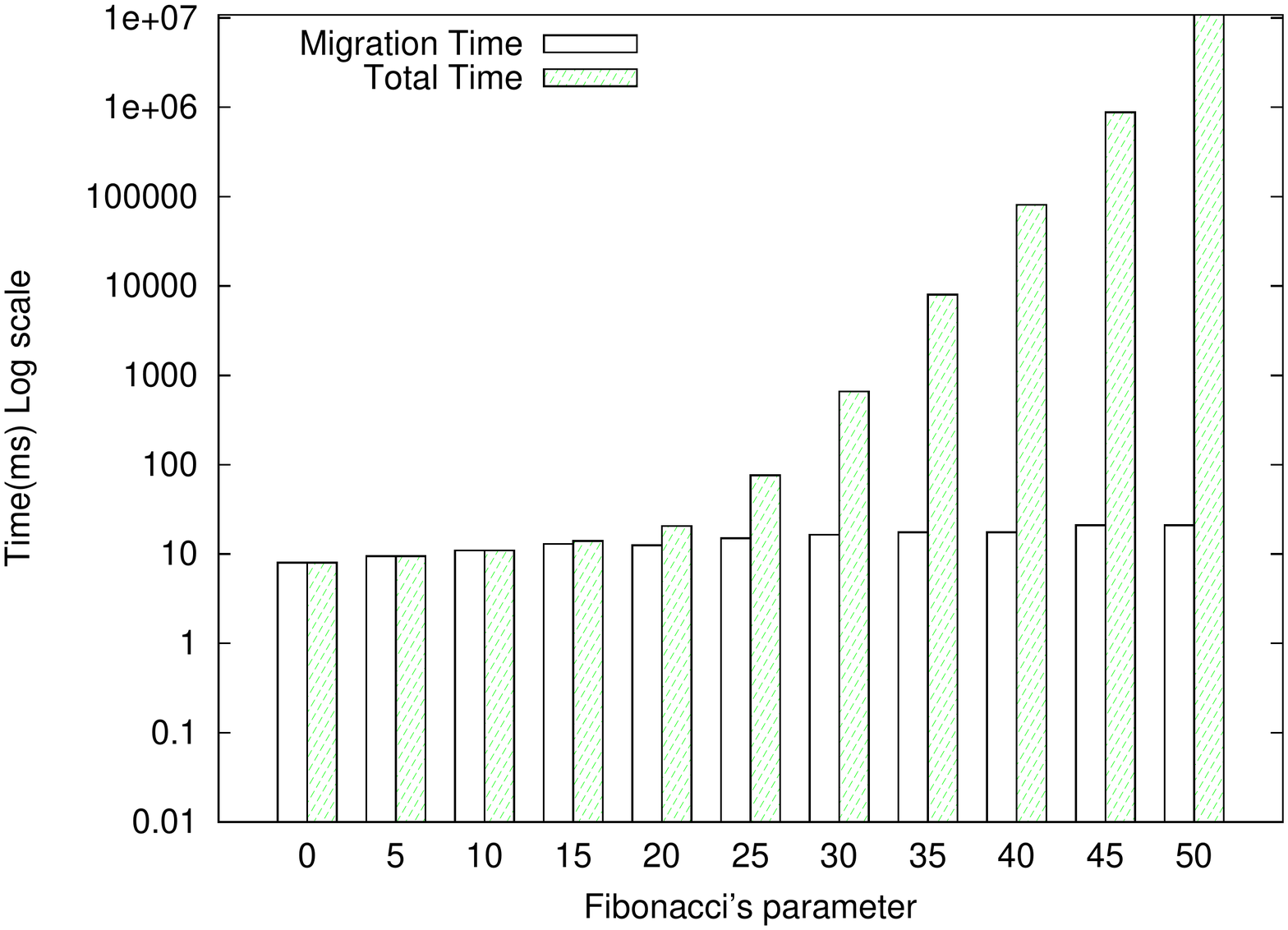}
\vspace{-0.5cm}
\caption{Migration of a Fibonacci execution}
\label{fig:FiboMigration}
\end{figure}

We included a suspension call that makes the program stop when 
the value of the parameter is 0. 
After the first suspension, a flag is set in order to allow the program to continue 
execution until the end. 
We have done this in order to measure the total execution time in presence of migration. 
(We could have also replaced function \textit{yield} for a dummy function on 
the representation before reinstalling it, with a similar effect).
The function was executed varying the Fibonacci's parameter value from 0 to 50 
at intervals of 5 units. 
The results are shown in Figure \ref{fig:FiboMigration}.

As the figure shows, the cost of the migration 
in relation to total execution time rapidly decreases 
with the increase of computation until it is almost negligible. 
Thus, the Fibonacci function is an example of a computation pattern 
that could take advantage of 
migration for \textit{Opportunistic Computing} or load balancing purposes, 
since the migration time is compensated by the computation time after a certain entry. 
We can also see that migration time grows steadily but very slowly. 
This is because every activation record has a reference to the same function 
(the Fibonacci function), which is reified only once, and thus the amount of new 
information with every new activation record remains constant and small.

\subsubsection{K-nearest neighbors algorithm (k-NN)}

The k-nearest neighbors algorithm (k-NN) is a method for classifying objects based on 
the closest examples in a training set, frequently used in data mining applications. 
An object is included in the most common class amongst its k-nearest neighbors, according 
to some measure of distance. 
In our implementation, we chose the cosine distance.

We implemented capturing and restoration of a coroutine executing the k-NN algorithm for 
classifying a relatively small database of 7 Mb (but the same procedure can be used for larger databases). 

In k-NN, the training table is traversed for every record of the test table, to compute 
the distance with the training records. 
We measured the total execution times of the application without and with migration. 
The experiment without migration consists in executing a Lua implementation of the k-NN algorithm. 
The experiment with migration consists in executing the program in a machine A until it reaches half the number of records in the test base, then it yields and the platform initiates the migration of the execution to a machine B. On arrival the platform at B transmits the state file back to A: 
\begin{enumerate}
\item Platform A captures, saves to disk and transmits from A to B the file containing the state, as well as the working files.
\item At B the plaform receives the file(s) and transmits it back to A. We calculate the transmission time as the time to transmit the files from A to B and back divided 2. 
\item The platform in A receives the file(s), loads it and restores the computation. That is, the file is openned and its content loaded into a buffer, then the values of the computation are recreated from the string loaded. 
\end{enumerate}
Finally, the restored coroutine is executed until its end. The transfer involved a buffer of 14581.992 kb. 

We performed 7 replications to guarantee an error of 0.9\% with a confidence interval of 95\%. 
The measured time for execution without migration was in media $446954.85ms$, while the time with migration reached $462505.78ms$. The total migration time (capture, save, transmission,load and restoration) amounts in media to $15051.35ms$. Since the program halted after $225945.42ms$, we can see that in that cenario (as for higher halting times) it is preferable to migrate instead of stopping the program and starting it from the beginning in other machine. 


\section{Final remarks}\label{sec:final}

This paper argued that
programming languages should offer mechanisms for fine-grained capture and 
restoration of the execution state in order to allow the implementation of different 
policies.
To illustrate this idea, we presented an API for reifying and installing computations. 
Our proposal is different from others in 
allowing the reification of the execution state of running computations in 
the form of fine-grained data structures that can be freely manipulated 
by the programmer.
With this API, it is possible to control factors that are typically predefined
in black-box serialization frameworks, 
such as granularity, the amount of execution state to be transferred 
or persisted, and the way the computation will be rebound to the new local context. 

We  have shown that  this approach allows implementing various and 
powerful functionalities. 
A possible drawback is that the programming burden augments, as 
does the chance of dealing with representation  inconsistencies. 
The idea is that an API such as luaNua be used in the development of libraries 
which implement different policies,
while still allowing direct access to the API for more specific applications.
The extended version of the LOOP library that we described is an
example of such a policy-implementing layer.

One of the insights we gained with our work is the
understanding the mechanisms that a general purpose language should provide 
in order to support heterogeneous capture and restoration of computations.
We believe the following set of mechanisms should be offered:
(i) fine-grained reification of computations, 
(ii) installation of composable computations,
(iii) binding of the installed computations to the new environment, and 
(iv) support for restarting the execution from a specific point. 
With these facilities, languages can provide indirect support
for persistency, checkpointing, and migration.

Future work includes exploring the flexibility we advocated,
building libraries with different policies for distributed systems based on Lua.
Besides their role in further evaluating our proposal,
these will also be used as support in other research, for instance in investigating
the management of concurrency levels in concurrent servers and in
opportunistic computing.

\bibliographystyle{elsarticle-harv}
\bibliography{main}
\end{document}